\documentclass[12pt]{article}
\usepackage{epsfig}
\def\Z {\rm {\bf Z}}
\def\R {\rm {\bf R}}
\def\1 {\rm {\bf 1}}
\begin{document}
\renewcommand{\arraystretch}{1.5}

\rightline{UG-7/97}
\rightline{hep-th/9707130}
\rightline{July 1997}
\vspace{1truecm}
\centerline{\bf GAUGED SUPERGRAVITY FROM DIMENSIONAL REDUCTION}
\vspace{1cm}
\centerline{\bf E.\ Bergshoeff \footnote{\tt e.bergshoe@phys.rug.nl},
M.\ de Roo \footnote{\tt m.de.roo@phys.rug.nl},
E.\ Eyras \footnote{\tt e.a.eyras@phys.rug.nl}}
\vspace{.4truecm}
\centerline{\it Institute for Theoretical Physics}
\centerline{\it Nijenborgh 4, 9747 AG Groningen}
\centerline{\it The Netherlands}
\vspace{2truecm}
\centerline{ABSTRACT}
\vspace{.5truecm}
We perform a generalised Scherk-Schwarz reduction of the 
effective action of the heterotic string on $T^6$ to obtain a 
massive $N=4$ supergravity theory in four dimensions. 
The local symmetry-group of the resulting $d=4$ theory includes a
Heisenberg group, which is a subgroup of 
the global $O(6,6+n)$ obtained in the standard reduction. 
We show explicitly that the same theory can be obtained by gauging
this Heisenberg group in $d=4, N=4$ supergravity.

\newpage

\noindent{\bf 1.\ Introduction}
\vspace{0.5cm}

\noindent Dimensional reduction is a key to understanding the interplay
 between various dualities in string theory. Many results in
 the field of string dualities have been obtained using standard 
 toroidal reductions. Recently there has been a renewed interest 
 \cite{BdRGPT, CLPST} in
 the generalisation of toroidal reduction introduced long ago
 by Scherk and Schwarz \cite{SS}. 
 Its basic property is that it allows a certain dependence of
 the fields on the coordinates which are wrapped around the torus.
 The result is usually that parameters with the dimension of mass
 are introduced into the resulting theory. 

A nontrivial feature of the Scherk-Schwarz reduction is that, although
before reduction some of the fields depend on the torus coordinates,
the reduced theory is independent of these coordinates. To achieve this
one specifies the particular dependence of the fields
on the torus coordinates by
using a global symmetry of the theory. In the original work of Scherk
and Schwarz, the global symmetry used for this purpose was a compact
subgroup of the internal symmetry group of the theory.
The Scherk-Schwarz mechanism in this form was applied 
to the six-index formulation of $d=10$ supergravity in \cite{Ch} and
to the effective action of the heterotic string on $T^6$ in \cite{TWPZ},
to obtain a gauged $d=4, N=4$ supergravity theory with a positive semidefinite 
potential; similar work on $d=11$ supergravity and $M$-theory was done in
\cite{CSS,DG}.

Almost coincident with the work of Scherk and Schwarz, 
it was realized that supergravity
theories have extra noncompact global symmetries \cite{CJ}. It is natural
to ask oneself whether one of these extra global symmetries can be used
as well in the Scherk-Schwarz mechanism. 
Recently, it has been shown that indeed this can be done in the reduction
of IIB supergravity to nine dimensions \cite{BdRGPT}.
The motivation of \cite{BdRGPT} was to establish duality rules relating
type IIB supergravity
to the massive version of type IIA supergravity \cite{Romans}.
The relevant noncompact symmetry used was a particular $SL(2,R)$ symmetry 
involving
 the Ramond-Ramond scalar of IIB supergravity. Basically the RR scalar
$\ell$ was replaced by $\ell' = \ell + my$, where
 $y$ is the coordinate over which one reduces. The fact that this
 global shift of the field $\ell$ is a 
 symmetry of the ten-dimensional action
 ensures that the linear $y$-dependence disappears in the reduction. The
 nine-dimensional theory then contains the massive parameter $m$.
 
The purpose of this letter is to show that the noncompact symmetry used
in \cite{BdRGPT} can also be applied in a more general 
context involving more scalars. To be explicit, we will start with
 the standard toroidal compactification of the heterotic string
 effective action on $T^5$ \cite{MS}. This theory has a global
 $O(5,5+n)$ symmetry\footnote{For the heterotic string $n=16$. It
 is convenient to keep $n$ arbitrary.}, which we use in a
 Scherk-Schwarz reduction to $d=4$ thereby giving a linear
 $x^4$-dependence to $n+8$ scalars. We find that
 the Scherk-Schwarz reduction induces a local nonabelian symmetry
 in the resulting $N=4, d=4$ supergravity theory. We establish that
 this generalised reduction is equivalent to the gauging of this
 nonabelian group in matter-coupled $N=4, d=4$ supergravity
 \cite{dRBKS}.

Let us first give an example of our use of the Scherk-Schwarz mechanism.
Consider a complex scalar field $\hat\lambda$ coupled to gravity in 
$d$ dimensions\footnote{This example is similar to the reduction of
the type $IIB$ string in \cite{BdRGPT}.}:
\begin{equation}
{\cal L}_d = \sqrt{|\hat g|} \bigg\{ {\hat R} + {{\partial {\hat \lambda} 
\partial \bar {\hat \lambda}}	
\over |{\hat \lambda} - \bar {\hat \lambda}|^2 } \bigg\} \, .
\end{equation}
\noindent $d$-dimensional fields are indicated by hats.
This lagrangian has a modular invariance $SL(2,\R)/\Z_2$:
\begin{equation}
{\hat \lambda} \rightarrow {c + d {\hat \lambda} \over a + b 
{\hat \lambda}}\, .
\end{equation}
\noindent For $b=0$ and $a=d=1$, we have  a 
$1$-parameter subgroup
isomorphic to $(\R, +)$. The real components 
of ${\hat \lambda}$ transform as
${\hat \lambda}_1 \rightarrow {\hat \lambda}_1 + c$ and
${\hat \lambda}_2 \rightarrow {\hat \lambda}_2$.
We use this subgroup in the generalised reduction.
The spacetime coordinates are divided as ${\hat x}_{(D)}=(x_{(D-1)},y)$ and
the generalized reduction rules for the scalars become
\begin{equation}
{\hat \lambda}_1 ({\hat x})= \lambda_1(x) + m y
\, , \,\,\,\, {\hat \lambda}_2 (\hat x) = \lambda_2(x) \, .
\end{equation}
\noindent This results in the following lagrangian in $d-1$ dimensions:
\begin{eqnarray}
\lefteqn{ {\cal L}_{d-1} = \sqrt{| g |} \bigg\{ R + {1 \over 2} 
(\partial \phi)^2 - { 1 \over 4}e^{ \sqrt{2(D-2) \over (D-3)} \phi} F^2 (A)}
\nonumber \\
&&+ {1 \over 2 \lambda_2^2} \lbrack (\partial \lambda_2)^2 +
( {\cal D} \lambda_1)^2 \rbrack - {1\over 4 \lambda_2^2}
m^2 e^{- \sqrt{ 2(D-2) \over (D-3)} \phi} 
\bigg\} \, .
\end{eqnarray}
\noindent The derivative ${\cal D}$ is defined as
${\cal D}_\mu \lambda_1 = \partial_\mu \lambda_1 - mA_\mu$.
The modular symmetry is broken down to a one-parameter local subgroup,
for which ${\cal D}$ is the covariant derivative\footnote{Throughout
this paper we will use $\Delta$ to indicate finite transformation,
defined for any field $F$ by 
$\Delta F = F' - F$, where $F'$ is the transformed field.
We will use $\delta$ for infinitesimal transformations.}:
\begin{equation}
\Delta {\lambda}_1 = m \beta (x), \, \qquad
\Delta A_\mu = \partial_\mu \beta (x) \, .
\end{equation}
\noindent The gauge field $A_\mu$ has become massive, as can be seen 
by going to the gauge $\lambda_1=0$.

In this example it is straightforward to see that this result  
 can also be obtained by first
 doing a standard reduction and by then 
 gauging the appropriate one-dimensional
 subgroup of $SL(2,\R)/\Z_2$ in $d-1$ dimensions. 
 Note however that by doing this one does not recover the 
 dilatonic potential. If this example is embedded in a supersymmetric
 theory, the potential can be recovered by requiring supersymmetry.

In the next section we will consider a similar, but more complicated
example. A basic difference with the above example will be that,
after reduction, three instead of one symmetries receive $m$-dependent
modifications. Correspondingly, this example involves the gauging of a 
three-dimensional group, which turns out to be the Heisenberg group.

\newpage
\noindent{\bf 2.\ Generalized reduction}
\vspace{0.5cm}

\noindent We carry out a standard reduction of the low energy effective 
 action of the heterotic string on $T^5$ to obtain the
 five dimensional $N=4$ supergravity theory. The action for the 
 bosonic fields is \cite{MS}\footnote{The hats 
 indicate five-dimensional
 fields and coordinates, the absence of a hat implies that the corresponding
 object is four-dimensional. We use the notation and conventions of
\cite{Behrndt}.}:
\begin{eqnarray}
S&=&  \int d^5x \sqrt {|\hat g|} e^{-2 \hat \phi}
\bigg\{\hat R - 4 (\partial \hat \phi)^2 - {1 \over 8}
{\rm tr}( \partial_{\hat \mu} \hat{\cal M} \partial^{\hat \mu} 
{\hat{\cal M}}^{ -1} ) \nonumber \\
&&+ {3 \over 4} {\hat H}_{\hat\mu\hat\nu\hat\rho}^2 +
 {1 \over 4} {\hat F}({\hat{\cal A}})_{\hat \mu \hat \nu}^i
 {\hat{\cal M}}^{ -1}_{ij} {\hat{F}}({\hat{\cal A}})^{\hat \mu \hat \nu j}
\bigg\} \, .
\label{eq:fivedimensionalaction}
\end{eqnarray}
\noindent This action has $O(5,5+n)$ symmetry \cite{MS}. 
 The scalars
 parametrize an $O(5,5+n)$ element ${\cal M}^{-1}$:
\begin{equation}
{\hat{\cal M}}^{-1} = \left ( \begin{array}{ccc}
 -{{\hat\kappa}_1}^2 - {{\hat{\ell}}^4  \over 4{{\hat\kappa}_1}^2}
         + {\hat\ell}^a {\hat{M}}_{ab}^{-1} {\hat\ell}^b  &  
  {1 \over 2{{\hat\kappa}_1}^2} {\hat\ell}^2 & 
  {\hat\ell}^a {\hat{M}}_{ab}^{-1} - { {\hat\ell}^2 
         \over 2{{\hat\kappa}_1}^2} {\hat\ell}^a {L}_{ab} \\ 
  {1 \over 2{{\hat\kappa}_1}^2} {\hat\ell}^2 & 
  - {1 \over {{\hat\kappa}_1}^2} &
  {1 \over {{\hat\kappa}_1}^2} {\hat\ell}^a {L}_{ab} \\
  {\hat{M}}_{ab}^{-1} {\hat\ell}^b - 
      {{\hat\ell}^2 \over 2{{\hat\kappa}_1}^2} {L}_{ab}{\hat\ell}^b &
  {1 \over {{\hat\kappa}_{1}^2}} {L}_{ab} {\hat\ell}^b &
  {\hat{M}}_{ab}^{-1} - {1 \over {{\hat\kappa}_1}^2} {L}_{ac} 
      {\hat\ell}^c {\hat\ell}^d {L}_{db}
\end{array} \right )  
\, .
\label{eq:fivedim-parametscalars}
\end{equation}
\noindent Here ${\hat M}^{-1}$ 
 is an element of $O(4,4+n)$, containing $4(4+n)$
 independent scalars, and $L$ is the invariant metric of $O(4,4+n)$.
 ${\hat\ell}^2$ equals $\ell^aL_{ab}\ell^b$.
 The indices have the range $i,j= 1,\ldots,10+n$, 
 $a,b = 1,\ldots,8+n$.

For the Scherk-Schwarz reduction we use the subgroup of $O(5,5+n)$
 under which the scalars ${\hat\ell}_a$ transform by constant shifts. 
 These transformations take on the form:
\begin{equation}
{\hat \omega}(\Lambda) = \left ( \begin{array}{ccc} 1 & 0 &       0    \\
			-{\Lambda^2 \over 2} & 1 & \Lambda^a L_{ab} \\
			-\Lambda^a & 0 & \delta_{ab} \end{array} \right )\, .
\label{transformation}
\end{equation}
\noindent Under these transformations $\hat{\cal M}\to {\hat \omega} 
 {\hat{\cal M}}{\hat \omega }^{T}$\footnote{Recall that 
 $\hat{\cal M}^{-1} = {\cal L}{\hat{\cal M}}{\cal L}$, where ${\cal L}$ is the
 $O(5,5+n)$ invariant metric.},
 which implies that ${\hat \ell}^a \to
 {\hat\ell}^a + \Lambda^a$, while $\hat M$ and ${\hat\kappa}_1$
 are invariant.
 The vector fields transform as ${\hat {\cal A}}_{\hat \mu} \to 
 \hat \omega 
 {\hat {\cal A}}_{\hat \mu}$. 
 Writing out ${\hat{\cal A}}$ in terms of $n+8$ vectors ${\hat V}^a$
 which transform as a vector under $O(4,4+n)$, and 
 vectors $A$ and $B$ which are Kaluza-Klein and winding vectors
 obtained in the reduction from six to five dimensions, we get
 the following transformations under $\hat\omega$:
\begin{eqnarray} 
{\hat\ell}^a &\to& {\hat\ell}^a + \Lambda^a\, , \nonumber \\
{\hat A}_{\hat \mu} &\to& {\hat A}_{\hat \mu}\, , \nonumber \\
{\hat B}_{\hat \mu}  &\to& {\hat B}_{\hat \mu} - {1 \over 2}{\Lambda^2}
{\hat A}_{\hat \mu} + \Lambda^a L_{ab} {\hat V}_{\hat \mu}^b\, ,  \\
{{ {\hat V}_{\hat \mu}}}^{a} &\to& {\hat V}_{\hat \mu}^a - \Lambda^a 
{\hat A}_{\hat \mu}\, .\nonumber
\label{fieldtransformation}
\end{eqnarray}

In the implementation of the Scherk-Schwarz reduction to four dimensions
 we include in the relation between four- and five-dimensional
 fields a dependence on the fifth coordinate $y$, by choosing $\Lambda$
 $y$-dependent. 
The generalized reduction rules, expressing the 5-dimensional
fields in terms of 4-dimensional ones, become
\begin{eqnarray}
&&{\hat g}_{\mu\nu} = g_{\mu\nu} - (\kappa_2)^2 A^{(2)}_\mu A^{(2)}_\nu\,,\quad
  {\hat g}_{\mu y}  = - (\kappa_2)^2 A^{(2)}_\mu\,,\nonumber\\
&&{\hat g}_{yy}     = -(\kappa_2)^2\,,\quad 
   \hat\phi = \phi + {1\over 2}\log\kappa_2\, , \nonumber\\
&&{\hat \ell}^a = \ell^a + \Lambda^a (y) \,,\quad
  {\hat\kappa}_1 = \kappa_1\,,\quad
  {\hat M} = M\,, \nonumber\\
&&{\hat A}_{\mu} = A_\mu + a A^{(2)}_{\mu} \,,\quad {\hat A}_y = a \, ,
\nonumber \\
&&
\label{generalizedreductionrules}
  {\hat B}_{\mu} = B_\mu - {1 \over 2}{\Lambda^2 (y)} A_\mu +
      \Lambda^a (y) L_{ab} V_{\mu}^b + A_{\mu}^{(2)} ( -{1 \over 2}
  {\Lambda^2 (y)} a + b + {\Lambda^a (y)}
      L_{ab} v^b) \, ,
\\
&&{\hat B}_{y} = b - {1 \over 2}{\Lambda^2 (y)} a
     + {\Lambda^a (y)} L_{ab} v^b \, ,
\nonumber \\
&&{{\hat V}_{\mu}}^a = V_{\mu}^a - {\Lambda^a(y)}
    A_\mu + A_{\mu}^{(2)} ( v^a - {\Lambda^a (y)}a) \,,\quad
  {\hat V}_{y}^a = v^a - \Lambda^a(y)a \, ,
\nonumber \\
&&{\hat B}_{y \mu} = B^{(2)}_\mu - {a \over 2}B_\mu - {b \over 2} A_\mu
	- {1 \over 2} v^a L_{ab} V_\mu^b \, , \nonumber \\
&&
{\hat B}_{\mu \nu} =
	B_{\mu \nu} +A^{(2)}_{[ \mu}B^{(2)}_{\nu]} 
	-aA^{(2)}_{[\mu}B_{\nu]} -bA^{(2)}_{[\mu}A_{\nu]} 
        -v^a L_{ab} A^{(2)}_{[\mu}V_{\nu]}^b \, .
\nonumber
\end{eqnarray}
\noindent In the reduction we will generate in the action terms proportional
 to  $\partial_y \Lambda^a$, at most to the second power, 
 so that the $y$-dependence disappears in four dimensions when
 $\Lambda^a=m^ay$.
The scalar kinetic term reduces as follows:
\begin{eqnarray}
&& {1 \over 8} {\rm tr}( \partial_{\hat \mu} {\hat {\cal M}} 
\partial^{\hat \mu} {\hat {\cal M}}^{-1}) =
{1 \over 8} {\rm tr}( \partial_\mu {M} 
\partial^\mu {M}^{-1})  
  - ( \partial {\rm ln} \kappa_1)^2 
 - {1 \over 2{\kappa_1}^2{\kappa_2}^2} m^a M_{ab}^{-1} m^b 
\nonumber \\
&&\quad + {1 \over 2{\kappa_1}^2} (\partial_\mu \ell^a - m^a A_{\mu}^{(2)})
     M_{ab}^{-1} (\partial^\mu \ell^b - m^b {A^{(2)}}^{\mu})\,.
\label{scalars}
\end{eqnarray}
\noindent As expected we find the local symmetry
$\Delta \ell^a = m^a \beta(x) , \,
\Delta A_{\mu}^{(2)} = \partial_{\mu} \beta(x) $,
 with $\beta (x)$ an arbitrary function of the
 four-dimensional coordinates.
 We also obtain a scalar potential that depends on the $O(4,4+n)$
 scalars and, in the full action, on the dilaton.

For the vectors we can make the $O(6,6+n)$ structure explicit, by making
 $m^a$-dependent modifications:
 Define
\begin{equation}
{\cal F}_{\mu \nu} ({\cal A}^I) \equiv 
\left ( \begin{array}{c} {\cal F}_{\mu \nu}^{(2)}(A) \\
{\cal F}_{\mu \nu}^{(2)}(B) \\
{\cal F}_{\mu \nu}(A) \\
{\cal F}_{\mu \nu}(B) \\
{\cal F}_{\mu \nu}(V^a) \end{array} \right ) 
= \left ( \begin{array}{c}
F_{\mu \nu}^{(2)} (A) \\
F_{\mu \nu}^{(2)} (B) + 2m^a L_{ab} A_{[ \mu} V_{ \nu ]}^b \\
F_{\mu \nu} (A) \\
F_{\mu \nu} (B) + 2 m^a L_{ab} V_{[ \mu}^b A_{\nu ]}^{(2)} \\
F_{\mu \nu} (V^a) + 2 m^a A_{[ \mu }^{(2)} A_{\nu ]} 
\end{array} \right ) 
\label{tensor-redefinition}
\end{equation}
\noindent  for $I=1,...,12+n$. All vector kinetic terms 
 can now be gathered  in the expression
\begin{equation}
 - {1 \over 4} {\cal F}_{\mu \nu} {\cal N}^{-1} {\cal F}^{\mu \nu}\,,
\label{vectorfields}
\end{equation}
\noindent  where ${\cal N}^{-1}$ is an $O(6,6+n)$ element parametrizing the
 $6(6+n)$ scalars in four dimensions:
\begin{equation}
{\cal N}^{-1} = \left ( \begin{array}{ccc}
-{\kappa_2}^2 - {z^4  \over 4{\kappa_2}^2}
+ {z}^i {\cal M}_{ij}^{-1} {z}^j  &  {1 \over 2{\kappa_2}^2} {z}^2 & 
{z}^k {\cal M}_{kj}^{-1} - { {z}^2 \over 2{\kappa_2}^2} {z}^k
{\cal L}_{kj} \\
 {{z}^2 \over 2{k_2}^2} & - {1 \over {\kappa_2}^2} &
 {1 \over {\kappa_2}^2} {z}^i {\cal L}_{ij} \\
{\cal M}_{ij}^{-1} {z}^j - {{z}^2 \over 2{\kappa_2}^2} {\cal L}_{ik}{z}^k &
{1 \over \kappa{_2}^2} {\cal L}_{ik} {z}^k &
{\cal M}_{ij}^{-1} - {1 \over {\kappa_2}^2} {\cal L}_{ik} 
{z}^k {z}^l {\cal L}_{lj} 
\end{array} \right ) \, .
\label{fourdim-parametscalars}
\end{equation}
\noindent Here  ${\cal M}^{-1}$ is the same matrix as
 (\ref{eq:fivedim-parametscalars}), but now without hats. The 
 $10+n$ scalars
 $z^i$ correspond to $a,b$ and $v^a$, $z^2$ is now $z^i{\cal L}_{ij}z^j$.

The 3-form strength tensor $H_{\mu \nu \rho}$ can be written as
\begin{equation}
H_{\mu \nu \rho} =
 \partial_{[\mu} B_{\nu \rho]} -
 m^a L_{ab} A_{[\mu}^{(2)} A_{\nu} V_{\rho ]}^b 
 + {1 \over 2} \eta_{IJ}{\cal A}_{[ \mu}^I
 {\cal F}_{\nu \rho ]}({\cal A}^J) \, ,
\end{equation}
\noindent with $\eta_{IJ}$ the invariant metric of $O(6,6+n)$. We will need 
 its explicit form later on:
\begin{equation}
  \eta_{IJ} = \left(\begin{array}{ccc}
           \sigma_1 & 0 & 0 \\
             0 & \sigma_1& 0 \\
             0 & 0  &  L_{ab} 
         \end{array}
         \right) \,.
\label{eta}
\end{equation}
The derivatives of the scalars $z^i$ can be  conveniently rewritten with
 $m^a$-dependent contributions:
\begin{equation}
{\cal D}_{\mu}z^i = \left ( \begin{array}{c} 
\partial_\mu a \nonumber \\
\partial_\mu b - m^a L_{ab}( V_{\mu}^b + v^b A_{\mu}^{(2)}) \nonumber \\
\partial_\mu v^a + m^a ( A_{\mu} + a A_{\mu}^{(2)}) 
\end{array} \right )\, ,
\label{covariant-scalars}
\end{equation}
\noindent The complete action then takes on the form:
\begin{eqnarray}
&& S_{4D, {\rm massive}} = \int d^4 x \sqrt{|g|}
 e^{-2 \phi} \bigg\{R - 4{(\partial \phi)}^2  - {1 \over 8} 
           {\rm tr} ( \partial_\mu {M} \partial^\mu {M}^{-1}) 
  + (\partial {\rm ln} \kappa_1)^2 + ( \partial {\rm ln} \kappa_2)^2
\nonumber \\
&&\quad
+ {1 \over 2{\kappa_1}^2{\kappa_2}^2} m^a M_{ab}^{-1} m^b 
- {1 \over 2{\kappa_1}^2} ( \partial^\mu \ell^a - m^a {A^{(2)}}^\mu)
M_{ab}^{-1}( \partial_\mu \ell^b - m^b A_{\mu}^{(2)}) \nonumber \\
&&\quad + {3 \over 4}H^2 +
  {1 \over 4} {\cal F}_{\mu \nu} {\cal N}^{-1} {\cal F}^{\mu \nu} 
- {1 \over 2{\kappa_2}^2} {\cal D}_{\mu}z {{\cal M}^{-1}}
{\cal D}^{\mu}z \bigg\} \, .
\label{massiveaction}
\end{eqnarray}

\vspace{0.5cm}
\noindent{\bf 3.\ Symmetries of the reduced action}
\vspace{0.5cm}
\label{symmetries}

\noindent The gauge transformations of the vector fields
 $A_\mu^{(2)},\ A_\mu$ and $V_\mu^a$,
 with parameters $\beta, \xi$ and $\eta^a$, respectively,
 that leave the action (\ref{massiveaction}) invariant
 obtain $m^a$-dependent modifications.
 To determine them, one should use the fact that the parameters 
 $\hat \xi, \hat \alpha$  and
 $\hat \eta^a$ corresponding to the gauge transformations of the
 vector fields $\hat A_\mu, \hat B_\mu$ and $\hat V_\mu^a$, respectively, are
 reduced as follows:
\begin{eqnarray}
\hat\xi &=& \xi\, ,     \, \nonumber\\
\hat\alpha &=& \alpha - {1\over 2}y^2m^aL_{ab}m^b\xi + ym^aL_{ab}\eta^b \, ,\\
\hat \eta^a &=& \eta^a - ym^a\xi     \, .\nonumber
\end{eqnarray}
Using this, we find that the $m^a$-modifications in the 
transformations, which we call $\Delta_\beta,\ 
 \Delta_\xi$ and $\Delta\eta$, respectively, take on the form:
\begin{eqnarray}
&&\label{Deltabetaxieta}
\begin{array}{rclrcl}
&&\Delta_\beta A_{\mu}^{(2)} = \partial_{\mu} \beta  \, ,&
&&\Delta_\beta B_{\mu}^{(2)} = 0 \, ,
\\
&&\Delta_\beta A_\mu = 0  \, ,&
&&\Delta_\beta B_\mu = - {{m^2 \over 2}} \beta^2 A_\mu + \beta m^a L_{ab} 
V_{\mu}^b \, ,
\\
&&\Delta_\beta V_{\mu}^a = -m^a \beta A_\mu \, ,&
&&\Delta_\beta B_{\mu \nu} = - B_{[ \mu}^{(2)} \partial_{\nu ]} \beta \, ,
\\
&&\Delta_\beta a = 0  \, ,& 
&&\Delta_\beta b = - {m^2 \over 2} \beta^2 a + \beta m^a L_{ab} v^b \, ,
\\
&&\Delta_\beta v^a = - m^a \beta a \, ,&
&&\Delta_\beta \ell^a =  m^a \beta \, . 
\\
&& \hfill&\hfill \\
&&\Delta_\xi  A_{\mu}^{(2)} = 0 \, ,&
  &&\Delta_\xi B_{\mu}^{(2)} = - {m^2 \over 2} \xi^2 A_{\mu}^{(2)} - \xi m^a 
     L_{ab} V_{\mu}^b \, ,
\\
&&\Delta_\xi A_\mu = \partial_\mu \xi \, ,&
&&\Delta_\xi B_\mu = 0  \, ,
\\
&&\Delta_\xi V_{\mu}^a = m^a \xi A_{\mu}^{(2)} \, ,&
&&\Delta_\xi B_{\mu \nu} = 
- A_{[\mu} \partial_{\nu ]} \alpha - B_{[ \mu} \partial_{\nu ]} \xi 
\, ,
\\
&&\Delta_\xi a = 0 \, ,&
&&\Delta_\xi b = 0 \, ,
\\
&&\Delta_\xi v^a = -m^a \xi \, ,&
&&\Delta_\xi \ell^a = 0 \, . \\
&&\hfill&\hfill\\
&&\Delta_\eta A_{\mu}^{(2)} = 0 \,, &
     &&\Delta_\eta B_{\mu}^{(2)} = m^a L_{ab} \eta^b A_\mu \, ,\\ 
&&\Delta_\eta A_\mu = 0 \, , &
      &&\Delta_\eta B_\mu = - m^a L_{ab} \eta^b A_{\mu}^{(2)} \, ,\\
&&\Delta_\eta V_{\mu}^a = \partial_{\mu} \eta^a \, ,&
&&\Delta_\eta B_{\mu \nu} = - L_{ab}V_{[ \mu}^a 
     \partial_{\nu ]} \eta^b \, ,\\
&&\Delta_\eta a = 0 \, ,& 
&&\Delta_\eta b = m^a L_{ab} \eta^b \, ,
\\
&&\Delta_\eta v^a = 0 \, ,& 
&&\Delta_\eta \ell^a = 0 \, .
\end{array}
\end{eqnarray}
\noindent Note that the $\Delta_\beta$ transformations are determined
 by performing a general coordinate transformation in five dimensions
 in the y-direction, with parameter $\beta(x)$.
 The remaining gauge transformations that do not obtain
 $m^a$-dependent
 modifications are given by:
\begin{eqnarray}
&&\Delta_\alpha B_\mu = \partial_\mu \alpha\,,\nonumber\\
&&\Delta_\gamma B^{(2)}_{\mu}= \partial_{\mu} \gamma\,,\\
&&\Delta_{\alpha,\gamma,\Sigma}B_{\mu\nu} = 
  \partial_{[\mu}\Sigma_{\nu]}
 - A_{[\mu}\partial_{\nu]}\alpha
  - A_{[\mu}^{(2)}\partial_{\nu]}\gamma\,. \nonumber
\end{eqnarray}
\noindent The generalised reduction has broken
 $O(6,6+n)$ invariance to an  $O(4,4+n)$ global symmetry, which
 acts in an obvious way on the four-dimensional fields.

The local infinitesimal transformations 
 $\delta_\beta$ , $\delta_\eta$ , $\delta_\xi$ ,
 $\delta_\gamma$ and $\delta_\alpha$ satisfy the following algebra:
\begin{equation}
\label{infalgebra}\begin{array}{ll}
\lbrack \delta_\beta , \delta_\xi \rbrack = \delta_{\eta^{\prime}} \, ,&
\eta^{\prime a} = m^a \beta \xi \, ,
\\
\lbrack \delta_\eta , \delta_\xi \rbrack = \delta_{\gamma^{\prime}} \, ,&
\gamma^{\prime} = - \xi m^a L_{ab} \eta^b \, ,
\\
\lbrack \delta_\beta , \delta_\eta \rbrack = \delta_{\alpha^{\prime}} \, ,&
\alpha^{\prime} = - \beta m^a L_{ab} \eta^b \, ,
\\
\lbrack \delta_\alpha , \delta_{any} \rbrack = 0 \, ,&
\lbrack \delta_\gamma , \delta_{any} \rbrack = 0 \, .
\\
\end{array}
\end{equation}
\noindent We see that the symmetries with parameters $\beta,\xi,\eta$
form an inhomogeneous 
 Heisenberg\footnote{The Heisenberg algebra is the algebra of 
 three operators $Q$, $P$ and $E$ that fulfill the canonical 
 conmutation relations $[Q,P]=E$, $[Q,E]=0$ and $[P,E]=0$.}
 algebra. The corresponding global transformations 
 have generators $T_\beta$, $T_\xi$ and $T^a_{\eta}$. The algebra is
\begin{equation}
   \lbrack T_\beta,T_\xi \rbrack = m^aL_{ab}T^b_\eta\,,\quad
   \lbrack T_\beta, T^a_\eta \rbrack = 0\,,\quad
   \lbrack T_\xi, T^a_\eta \rbrack = 0\,.
\end{equation}
Note that the only nontrivial commutation relation in this algebra involves
 the combination $m^aL_{ab}T^b$. This can also be seen from the
 $m^a$-dependent terms in the transformations (\ref{Deltabetaxieta}),
 which always contain the combinations $m^aL_{ab}V_\mu^b$ and
 $m^aL_{ab}\eta^b$. Therefore, effectively the nonabelian group which 
emerges from
 the Scherk-Schwarz reduction is three-dimensional.

\vspace{0.5cm}
\noindent{\bf 4.\ Equivalence with a gauged symmetry}
\vspace{0.5cm}

\noindent We will now show that (\ref{massiveaction}) can also be obtained by 
 starting from $m^a\equiv0$ (the standard reduction) and by gauging
 an appropriate subgroup of $O(6,6+n)$. This will give all terms in 
 (\ref{massiveaction}), except the scalar potential, which will be obtained 
 by supersymmetry.

Using indices $I,J=1,\ldots,12+n$ as in (\ref{tensor-redefinition}),
 and by considering the algebra (\ref{infalgebra}) we obtain
 structure constants $f_{IJ}{}^K$:
\begin{eqnarray}
{f_{3b}}^2 &=& m^a L_{ab} \, , \nonumber \\
{f_{b1}}^4 &=& m^a L_{ab} \, , \label{structure} \\
{f_{13}}^a &=& m^a   \, . \nonumber
\end{eqnarray}
\noindent Here the directions 1,2,3,4 and $a$ correspond to the symmetries
with parameters $\beta,\gamma, \xi,\alpha$ and $\eta^a$, respectively.
Assuming local transformations of the gauge fields of the
 form $\delta {\cal A}_\mu^I = \partial_\mu \lambda^I + 
 {f_{JK}}^I \lambda^J {\cal A}_\mu^K $ we recover the modified 
 field-strengths (\ref{tensor-redefinition}). Since the transformations
 of the scalars under global $O(6,6+n)$ transformations are
 given, we are uniquely led to covariant scalar derivatives in 
 (\ref{massiveaction}). 

By gauging the bosonic theory (\ref{massiveaction}) for $m^a=0$ we
 do not recover the scalar potential in (\ref{massiveaction}), since
 it is gauge invariant by itself. To obtain it we must use the fact
 that (\ref{massiveaction}) is the bosonic part of a supersymmetric theory,
 namely some version of $N=4$ supergravity. We should therefore be able 
 to obtain (\ref{massiveaction}), 
 now including the potential, from the results of \cite{dRBKS}.

In matter coupled $N=4$ supergravity the scalars parametrize an
 $O(6,6+n)/(O(6)\times O(6+n)) \times SU(1,1)/U(1)$ coset.
 The $SU(1,1)/U(1)$ coset corresponds to the dilaton and to the
 dual of $B_{\mu\nu}$. The scalars of $O(6,6+n)/(O(6)\times O(6+n)$
 can be  expressed in terms of real fields 
 $Z^{I}_a$, where $I=1,\ldots,12+n$ and $a=1,\ldots,6$, satisfying
\begin{equation}
  Z^{I}_a\eta_{IJ} Z^{J}_b = -\delta_{ab}\,. \label{constraint}
\end{equation}
\noindent For our purposes it is convenient to introduce the 
 combination\footnote{These variables were introduced in \cite{Wagemans}.}
 $Z^{IJ}\equiv Z^{I}_aZ^{J}_a$. The lagrangian of gauged
 $N=4$ supergravity can be expressed in terms of the $Z^{IJ}$. We give
 only the scalar kinetic terms and the potential, and to compare
 to (\ref{massiveaction}) we will use the string frame:
\begin{eqnarray}
 S_{4D,\ N=4} &=& {1 \over 2}\int d^4 x \sqrt{|g|}
   e^{-2 \phi} \bigg\{R - 4{(\partial \phi)}^2  
  + (\eta_{RS} +\eta_{RT}\eta_{SU}Z^{TU}){\cal D}_\mu Z_a^R {\cal D}^\mu
  Z^S_a \nonumber \\
 &&
\label{N=4action1}
\ + Z^{RU}Z^{SV}(\eta^{TW} + {2\over 3}Z^{TW}) f_{RST}f_{UVW}
  \bigg\}\,.
\end{eqnarray}
\noindent The structure constants are defined as $f_{RST}=f_{RS}{}^U\eta_{TU}$.
We must now show that these terms are equivalent to the corresponding terms 
in (\ref{massiveaction}) for the case of the group (\ref{structure}).
First of all note that (\ref{constraint}) implies that
$W^{RS} = \eta^{RS} + 2Z^{RS}$ is an element of
$O(6,6+n)$:
\begin{equation}
  W^{RS}\eta_{ST}W^{TU} = \eta^{RU}\,.
\end{equation}
Introducing $W$ in (\ref{N=4action1}) we obtain
\begin{eqnarray}
 S_{4D,\ N=4} &=& {1 \over 2}\int d^4 x \sqrt{|g|}
   e^{-2 \phi} \bigg\{R - 4{(\partial \phi)}^2  
  -{1\over 8} {\cal D}_{\mu} W^{RS} {\cal D}^{\mu} W_{SR}
 \nonumber\\
 &&
\label{N=4action2}
 + {1\over 12}\big\{2\eta^{RU}\eta^{SV}\eta^{TW} 
       + 3\eta^{RU}\eta^{SV}W^{TW} + W^{RU}W^{SV}W^{TW}\big\}
    f_{RST}f_{UVW}
  \bigg\}\,.
\end{eqnarray}
\noindent We can now identify $W^{RS} = {\cal N}_{RS}$, 
 which gives the correct kinetic term.
 In the potential we use that the only nonzero structure constants
 with lower indices are $f_{a13} = m^bL_{ba}$. Using the explicit
 form of ${\cal N}$ (\ref{fourdim-parametscalars}) 
 and $\eta$ (\ref{eta})
 we recover exactly the potential of (\ref{massiveaction}).
 
This shows that our massive theory
 (\ref{massiveaction}), obtained by the Scherk-Schwarz method,
 is equivalent to a gauged $N=4$ supergravity theory. 

\vspace{0.5cm}
\noindent{\bf 5.\ Conclusions}
\vspace{0.5cm}

\noindent In this letter we perform a generalised Scherk-Schwarz reduction
of the effective action of the heterotic string on $T^6$ to obtain a
massive supergravity theory in four dimensions. 
Our implementation of the Scherk-Schwarz method uses a noncompact
subgroup of $O(5,5+n)$ in the step from five to four dimensions.
We have shown explicitly that this theory can be obtained by
gauging a Heisenberg subgroup of the global $O(6,6+n)$ symmetry group
of $d=4,\ N=4$ matter coupled supergravity.

In the Einstein frame, with the
 dilaton kinetic term normalised to $1/2(\partial\phi)^2$,
 the scalar potential in (\ref{massiveaction}) 
 appears with a factor $\exp{(\phi)}$,  which would be
 $\exp\,(\sqrt{2/(d-2)}\,\phi)$ for a similar analysis in $d$ dimensions.
 The presence of such a scalar potential, with the
 same dilaton dependence, had to be
 assumed in \cite{MaSi} in order to obtain maximally symmetric black
 hole solutions in a $d=5$ context. It is interesting to note that
 the Scherk-Schwarz procedure, as well as gauging supergravity (see 
 \cite{TN} for an example in $d=7$), generates such potentials.
 In this respect it would be interesting to elucidate further the relationship
 between the Scherk-Schwarz procedure and stringy reduction 
 methods\footnote{For recent work on this, see \cite{ST} and
 references therein.}.
 
Clearly, by breaking the global $O(4,4+n)$ symmetry in five dimensions,
there are further possible shift symmetries that can be used in the 
dimensional reduction. A study of these possibilities has been made in 
\cite{CLPST}. It would be interesting to see whether they also lead
to gauged supergravities in four dimensions and to determine the
corresponding gauge group. In this way, a whole class of gauged
supergravities could be given a higher-dimensional interpretation. 
We do not expect that this can be done for all gauged
supergravity theories. For instance, it cannot be done for the
massive $d=10$ supergravity theory of Romans (other exceptions
have been given in \cite{CLPST}). It would be interesting to
see whether these exceptional cases can be given a higher-dimensional
interpretation by some other techniques.

\vspace{0.5cm}
\noindent{\bf Acknowledgements}
\vspace{0.5cm}

\noindent E.E.\  thanks H.\ Boonstra, B.\ Janssen and J.~Maharana for
useful remarks.
This work is part of the Research program of the
``Stichting voor Fundamenteel Onderzoek der Materie''(FOM).
It is also supported by the European Commission TMR programme
ERBFMRX-CT96-0045, in which E.B. and M. de R. are associated
to the University of Utrecht.


\end{document}